\documentclass[acmlarge]{acmart}
\AtBeginDocument{%
  }

\usepackage{color,colortbl}
\definecolor{Gray}{gray}{0.9}

\setcopyright{acmlicensed}
\copyrightyear{2025}
\acmYear{2025}
\acmDOI{XXXXXXX.XXXXXXX}

\acmJournal{POMACS}
\acmVolume{0}
\acmNumber{0}
\acmMonth{0}

\begin{document}

\title{A Use Case Lens on Digital Cultural Heritage}

%Other suggestions
% Bridging Infrastructures and Communities: A Framework for Robust Use Cases in Digital Cultural Heritage
%Method Matters: A Novel Framework for Capturing Use Cases in Digital Cultural Heritage
%Use Cases Reimagined: A Systematic Approach for Digital Cultural Heritage in eInfrastructures
% Mahendra: A new approach to creating use cases in Digital Cultural Heritage: working with einfrastructures
% Steven: Towards a Structured Approach to Designing Use Cases with Jupyter Notebooks [in a Federated Infrastructure] for Digital Cultural Heritage.

\author{Gustavo Candela}
\email{gcandela@ua.es}
\orcid{0000-0001-6122-0777}
%\author{Milena Dobreva}
%\email{webmaster@marysville-ohio.com}
\affiliation{%
  \institution{University of Alicante}
  \city{Alicante}
  \country{Spain}
}

\author{Milena Dobreva}
\orcid{0000-0002-2579-7541}
\affiliation{%
  \institution{University of Strathclyde}
  \city{Glasgow}
  \country{Scotland}}
\email{milena.dobreva@strath.ac.uk}

\author{Henk Alkemade}
\orcid{0000-0002-4413-5463}
\affiliation{%
  \institution{4DLandscapes}
  \city{The Hague}
  \country{The Netherlands}
}

\author{Olga Holownia}
\orcid{0000-0003-1800-6526}
\affiliation{%
  \institution{Internet Preservation Consortium}
  \city{Washington} 
  \state{District of Columbia} 
  \country{USA}
}
\email{olga@netpreserve.org}

\author{Mahendra Mahey}
\orcid{0000-0001-7608-6152}
\affiliation{%
  \institution{University of Strathclyde}
  \city{Glasgow}
  \country{Scotland}}
\email{mahendra.mahey@strath.ac.uk}

\author{Sarah Ames}
\orcid{0000-0002-0118-189X}
\affiliation{%
  \institution{National Library of Scotland}
  \city{Edinburgh}
  \country{Scotland}}
\email{s.ames@nls.uk}

\author{Karen Renaud}
\orcid{0000-0002-7187-6531}
\affiliation{%
  \institution{University of Strathclyde}
  \city{Glasgow}
  \country{Scotland}}
\email{karen.renaud@strath.ac.uk}

\author{Ines Vodopivec}
\orcid{0000-0002-3674-8630}
\affiliation{%
  \institution{AI4LAM}
  \city{-}
  \country{-}
  }
\email{ines.vodopivec@nb.no}

\author{Benjamin Charles Germain Lee}
\orcid{0000-0002-1677-6386}
\affiliation{%
  \institution{University of Washington}
  \city{Washington}
  \country{United States}}
\email{bcgl@uw.edu}

\author{Thomas Padilla}
\orcid{0000-0002-6743-6592}
\affiliation{%
  \institution{Bristlecone Strategy}
  \city{-}
  \country{United States}}
\email{thomas@bristleconestrategy.com}

\author{Steven Claeyssens}
\orcid{0000-0003-1110-5935}
\affiliation{%
  \institution{KB, National Library of the Netherlands}
  \city{The Hague}
  \country{The Netherlands}}
\email{steven.claeyssens@kb.nl}

\author{Isto Huvila}
\orcid{0000-0001-9196-2106}
\affiliation{%
  \institution{Department of ALM, Uppsala University}
  \city{Uppsala}
  \country{Sweden}}
\email{isto.huvila@abm.uu.se}

\author{Beth Knazook}
\orcid{0000-0003-3108-3921}
\affiliation{%
  \institution{Digital Repository of Ireland, Royal Irish Academy}
  \city{Dublin}
  \country{Ireland}}
\email{b.knazook@ria.ie}

\renewcommand{\shortauthors}{Candela et al.}

\begin{abstract}

This article proposes a novel methodological approach for developing use cases for CH e-infrastuctures documented using Jupyter Notebooks (JNs), enabling transparency and reproducibility. We also address the present problem of use cases that are not consistently documented to cover all key aspects that are derived from the use case literature review outside of CH field to define a useful use case. 

\textbf{Purpose.} Our primary objective is to explore the practices around creating and analysing use cases related to digital cultural heritage. Our review of the literature showed a substantial deviation in the depth and coverage of use cases and revealed the need for a more robust and consistent approach to creating use cases in a digital heritage context. We developed a framework to develop use cases to support the ongoing efforts to expand the use of eInfrastructures in the digital heritage domain as a first step. 
    
\textbf{Design/methodology/approach.} Our research design combines desk research of existing literature and analysing examples of use cases documented in projects. We examine the challenges and inconsistencies in the current practice of use case production in digital heritage. Finally, we synthesize a systematic process to generate use cases which is illustrated by five example use cases within the context.
    
\textbf{Findings.} We discovered that the current practice of use case design and creation in the domain of digital cultural heritage is highly inconsistent. Our analysis of issues enabled us to propose a more systematic approach to developing use cases in digital heritage, which we had approbated by developing five use cases which use eInfrastructures.
    
\textbf{Originality.} Our paper offers an original methodology for capturing use cases in digital cultural heritage that could be applied to different contexts.
    
\textbf{Research limitations/implications.} We focused on the digital heritage domain, reflecting on some of the specific communication needs of users within this area. Our findings and methodology may be applicable to other domains, but we have not explored these. More specifically, our work draws on developing five specific use cases involving user engagement with a specific eInfrastructure. Whilst we argue that these cases provide useful insights in use case development in general, they are not generalisable to cover all uses of federated eInfrastructures.
    
\textbf{Practical implications.} The practical implications of generating improved quality and consistency of use cases in digital heritage are substantial. Use cases serve as tools: (i) to describe the behaviour of systems, in this case, eInfrastructures under development, such as the European Cloud for Heritage Open Science (ECCCH) and the European Data Space for Cultural Heritage (DS4CH); (ii) to define the processes of communication with users and HPCs; (iii) to keep the communities behind eInfrastructures aligned with their aspirations. We developed this framework to empower practical developments in modern eInfrastructures through a more consistent use case creation practice.
    
\textbf{Social implications.} Users of eInfrastructures in digital heritage are diverse, they include researchers, educators, general citizens, and other types or categories. By offering improved means to capture and represent an increasing diversity of user perspectives to using eInfrastructures, our work contributes to the wider use and reuse of digital content in society. 

Our work impacts directly such infrastructures and communities as the International GLAM Labs Community, AI for Libraries, Archives, and Museums (AI4LAM) and Time Machine Organisation. This work advances the use of data research infrastructures within communities of researchers, scholars, students, GLAM (Galleries, Libraries, Archives, and Museums) institutions, and Cultural Heritage and Cultural and Creative Industries (CCIs).  

\end{abstract}




\received{}
\received[revised]{}
\received[accepted]{}

\maketitle

\section{Introduction}
GLAM (Galleries, Libraries, Archives, and Museums) and other Cultural Heritage (CH) institutions are exploring innovative ways to make their content available in the form of data \cite{tasovac:hal-02961317,DBLP:journals/jasis/Candela23}. They host and provide access to broad and diverse material that includes, but is not limited to, newspapers, webpages, born-digital files, posters, letters, postcards, bibliographic data, artworks, historical maps, and other objects relevant to the historical, cultural, and scientific heritage. These are captured in variety of formats: images, text, 3D models, audio and video. Collections are often multimodal, adding layers of complexity to management of access, use, and reuse.

In addition to a variety of digital cultural heritage content, there is an additional dimension of diversity: the \textit{users} of this content and how exactly they benefit from it. There are multiple potential contexts of use for different groups of actors including researchers, educators, creative industry professionals, and the general public, to name a few. Over the last few decades, the exploration of users and their needs has attracted significant attention, but also thorough how difficult they describe a comprehensive set of study at least in some comprehensiveness the majority degree of collections use. The needs conceivable of researchers have been explored in the literature both in general, and directly by multiple major eInfrastructure communities, including the Europeana Network Association (ENA) ~\cite{europeana, walsh_2016, av_2017} and AI4LAM~\cite{ai4am-use-cases}. The research shows that the diversity of research practices make it extremely difficult to describe a thorough set of scenarios of collections use that would match at least in some degree of comprehensiveness the majority of conceivable use cases. Still naturalistic use cases are a key component for establishing interoperable Digital Commons for both reproducible research and effective collaborative analysis of collections, including those of digital texts, historical data and maps, images, and digital twins.

It is still crucial that research is done in unexpected ways but the consolidation of digital research resources, including data and tools, has led to formation of identifiable patterns in how major communities of researchers approach to digital collections. Although approaching the colossal problem of matching resources with users through catering for these communities does not mean that all needs would be covered, addressing them can be a crucial step from addressing impractically niched or generic needs to making a real difference. This is due to the combined impact of two contemporary trends: (1) the transition from aggregated to federated resources and (2) the expanded use of Artificial Intelligence (AI) and Machine Learning (ML) tools.  

The most recent developments in eInfrastructures, including the European Collaborative Cloud for Cultural Heritage (ECCCH) and the European Data Space for Cultural Heritage (DS4CH), mark a major change in the approach to organising and accessing digital heritage and to its use and reuse. Both eInfrastructures implement federated access to collections and offer the possibility of more intense use of computationally ``intelligent'' tools. Within recent developments, the ECCCH is a shared and collaborative platform designed to provide heritage professionals and researchers with federated access to data, scientific resources, training, and advanced digital tools. This approach holds great promise for businesses, public administrations and individuals for creating environments that offer users flexibility and adaptability to cater for their diverse behaviours and needs. In parallel, DS4CH works to provide access to cultural heritage data from across Europe in a federated environment that promises sovereignty, trust, and security~\cite{DBLP:conf/icadl/DobrevaSI22, DBLP:conf/dolap/0001CR25}. In addition, the Time Machine Organisation (TMO) is a related international initiative for cooperation in technology, science and cultural heritage, focusing on digital twins, a virtual replica of a physical asset, process or system.
 
Common to all these European initiatives is that they aim at: (i) fostering collaboration between researchers, scholars, and communities to strengthen digital autonomy and sovereignty \cite{DBLP:conf/dolap/0001CR25}; (ii) promoting best practices for open and transparent research such as FAIR~(Findable, Accessible, Interoperable and Reusable) \cite{wilkinson2016fair,10.1108/GKMC-06-2023-0195}, Open Science as an approach to publishing and reusing data and code owned, developed and maintained by diverse communities \cite{Krewer2024Digital}; (iii) making the European Union (EU) a world class hub for human-centric and trustworthy AI~\cite{ai-eu}; and (iv) expanding public participation in research and innovation and thereby enhancing public trust in science \cite{ai-oe}. 

In addition, several communities and initiatives, such as the International GLAM Labs Community, Collections as Data, and AI4LAM have emerged during the last years to foster their reuse and support responsible and computational access \cite{open_glam_lab,padilla_thomas_2019_3152935}. Advances in technology, in particular AI and ML, have provided a new and novel context for researchers and scholars in terms of the availability of cloud services, data infrastructures, and unprecedented access to enormous data archives.

During recent years, Jupyter Notebooks (JNs) have emerged as an efficient and valuable instrument that facilitates reproducibility, community sharing, and Open Science \cite{DBLP:journals/jasis/CandelaCS23,sherratt_2025_15597489}. Within the range of ongoing efforts supporting the ECCCH, there are multiple examples of work in specific sectors of cultural heritage (e.g., music, textiles, archaeological data) and advanced analytical and ingestion tools \cite{eccch-previous}. However, a focused effort on harnessing the potential of JNs within ECCCH is still lacking, including how they can facilitate data integration of multiple collections and external sources of data such as Wikidata in Digital Humanities research. We note that in the ECCCH context, adding JNs offer opportunities to interactive creation and sharing of documents with live code, equations, visualisations, and documentation, currently lacking from the available tools. Our work aims to address this gap. 

Compared to previous research on users of eInfrastructures, the aim of this article is to describe a new methodology to advance the state-of-the-art for generating use cases for digital cultural heritage research. This work aims to answer the following research questions: 

(\textbf{RQ1}) What limitations and strengths there are in the previous practice on the creation of use cases in digital cultural heritage in general and within new research data infrastructures such as the ECCCH and DS4CH in particular? 

(\textbf{RQ2}) How can we define and assess a methodology to create use cases, focused on JNs, supporting the goal of an integrated, interoperable Digital Commons and reproducibility in Digital Humanities research?

Our main contributions are: (a) a methodology to define and implement use cases from the digital cultural heritage domain, and (b) the creation of five use cases based on this methodology. These contributions are intended to leverage the adoption and use of eInfrastructures, such as the ECCCH, by GLAM institutions and Cultural Heritage and Cultural and Creative Industries (CCIs), providing detailed documentation, training materials, and reproducible code.

\begin{figure}[ht]
    \centering
    \includegraphics[width=0.8\linewidth]{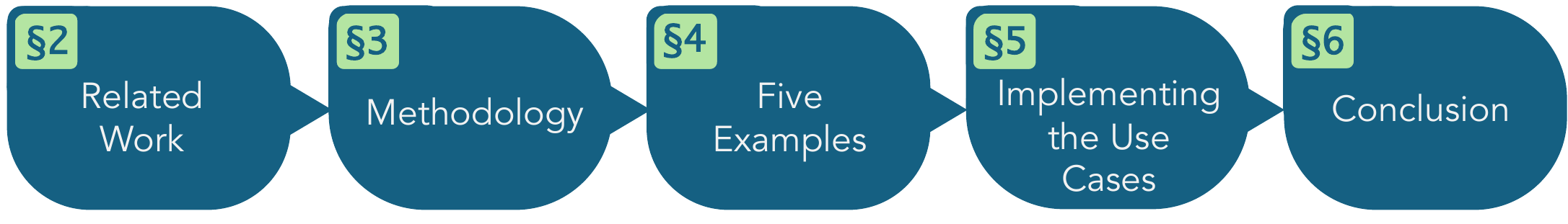}
    \caption{Paper Structure}
    \Description{Paper Structure}
    \label{fig:flow}
\end{figure}

As shown in Figure \ref{fig:flow}, our argument is structured as follows: After a brief overview of the current situation in digital cultural heritage, we introduce the methodology used to define and create use cases using engagement with ECCCH as an example. However, the methodology can be applied to other federated eInfrastructures. The article concludes with a brief summary of the proposed methodology and future work.

\section{Related Work}
\label{sec:sota}

GLAM and other CH institutions have been making digital collections available in a wide diversity of content and formats \cite{Disli_2025}. However, most of these datasets exist in data silos, hindering collaboration and resulting in duplicated efforts of data reuse and publication, inconsistencies, and missed opportunities in terms of research.

Several community-led initiatives have recently embarked on: (i) promoting the publication of digital collections supporting computational use such as Collections as Data~\cite{padilla_thomas_2019_3152935}; (ii) encouraging an innovative and responsible use of the digital collections following a reproducible approach based on the use of JNs such as the International GLAM Labs Community{\footnote{\url{https://glamlabs.io/}}} \cite{DBLP:journals/jasis/CandelaCS23,sherratt_2025_15597489}; (iii) providing detailed documentation of the digital collections using machine-readable formats and covering a wide diversity of metadata categories (e.g., biases, provenance and curation) such as Datasheets for Digital Cultural Heritage Datasets \cite{Alkemade-2023}; and (iv) applying AI to digital cultural heritage to gain insights using a broad range of methods such as Computer Vision, Large Language Models (LLM) and data enrichment~\cite{Candela-2025,DBLP:journals/jocch/MountantonakisT25}. Other platforms, such as Hugging Face,\footnote{\url{https://huggingface.co/}} have recently emerged with hard-working communities that promote the use of AI providing access to a wide range of cloud-based resources such as datasets, documentation, code and pre-trained models. 

The ECCCH is a shared and collaborative platform designed to provide heritage professionals and researchers with access to data, scientific resources, training, and advanced digital tools tailored to suit their needs. This platform is developed by the European Cloud for Heritage Open Science Project (ECHOES),\footnote{\url{https://www.echoes-eccch.eu/}} a project funded by the European Commission and UK Research and Innovation (UKRI) that brings together fragmented communities from the CH field such as Digital Humanities and Computer Science into a new community around the Digital Commons.

JN is an open source browser-based tool for creating interactive notebooks that combine text and code, document research workflows, code, data, and visualisations. It is ideal for interactive data science and scientific computing across disciplines, supporting programming languages including Python or R. JNs have been found to reduce many barriers to reproducibility and Open Science \cite{DBLP:journals/ese/PimentelMBF21}. They are designed to support reproducible research by enabling scientists to easily integrate computational narratives, combining text, code, charts, and documentation. Aligned with the ECCCH principles, JNs provide means to promote inclusion and democratise open and transparent access to knowledge and digital collections serving public good by and for all. In parallel, JNs introduce, however, new challenges relating to sustainability, AI integration and future reuse due to problems related to data access and leakages, software quality, bad programming practices, unresolved dependencies and platform differences in publicly available notebooks \cite{DBLP:journals/jasis/CandelaCS23,10.1145/3641539,10.1093/gigascience/giad113}. They require specialised knowledge for data repositories to curate and preserve \cite{bouquin}, and the introduction of AI adds further complexity to the work of maintaining them as functional, accessible resources. In humanities research, there is a growing interest among researchers and a growing supply by cultural heritage institutions, combined with recognition of challenges with size of collections and in measuring their impact \cite{doi:10.1177/03400352211065484}. JNs are a powerful tool to capture and represent the entire computational part of a research workflow and provide means to document other things as well such as code, metadata and dependencies. Table \ref{tab:jns} shows examples of JNs projects provided by GLAM institutions and research projects, and cloud services based on the employment of JNs. 

\begin{table}[ht]
    \centering
     \caption{Examples of collections and cloud services based on JNs. Note that all these examples are specific to digital cultural heritage}
    \label{tab:jns}
    \begin{tabular}{p{11cm}|p{1cm}}
        \toprule
        \textbf{Description} & \textbf{Source} \\
        \midrule
        National Library of Scotland & \cite{doi:10.1177/01655515231174386}\\
        \rowcolor{Gray}
        GLAM Jupyter Notebooks & \cite{DBLP:journals/jis/RomeroSES22} \\
       
        GLAM Workbench & \cite{sherratt_2025_15597489}\\
       \rowcolor{Gray}
        Library of Congress Data Exploration tutorials & \cite{loc-nb}\\
      
        Best practices and guidelines & \cite{DBLP:journals/jasis/CandelaCS23}\\
     \rowcolor{Gray}
        European Open Science Cloud (EOSC) - Interactive Notebooks & \cite{eosc-jns}\\
     
        Using Jupyter Notebooks to Process the Europeana Newspaper Text Resources & \cite{europeana-jns}\\
   \rowcolor{Gray}
        Turning JNs into data-driven web apps for different purposes & \cite{CLARKE2021100213,DBLP:journals/jocch/Candela25}\\
        
        Using JNs to run federated queries & \cite{disli_2025_15349480}\\
    \rowcolor{Gray}
        Extracting Collections as Data & \cite{gustavo_candela_2025_16747522}\\
        \bottomrule
    \end{tabular}
   
\end{table}

Within the range of ongoing efforts supporting the ECCCH, there are developments related to specific parts of the CH sector (e.g., music, textile heritage, archaeological data) and advanced analytical and ingestion tools. A range of new projects funded by the European Commission covers a wide diversity of content, including high-fidelity 3D models, 4D, digital twins, as bibliographical and stratigraphic data. They also provide a set of workflows that can be applied and assessed through use cases and pilots. These projects are planning to explore integration with several initiatives such as Europeana, EOSC, DS4CH, and Wikidata, among others. Innovations include, for example, an open library of Semantic Web and AI-based tools for multilingual metadata enrichment, the employment of explainable AI, the development of reusable digital twin standards, AI-powered bibliographic annotation, interoperable archival systems, and the development of predictive and simulation-based tools to support data-informed restoration decisions \cite{eccch-previous}. In general, they all promote and apply, to some extent, the FAIR principles and the long-term preservation of CH in Europe.  

Based on previous research conducted by multiple communities, this proposal is aligned with previous ECCCH-funded projects~\cite{eccch-previous} and complements them in terms of FAIR adoption, integration of both cross-disciplinary and domain-specific platforms and services such as EOSC, DS4CH, Wikidata and Wikimedia software~\cite{DBLP:conf/ercimdl/CandelaCHGDM24}, reuse of content and topics covered, contributing to the goal of a Digital Commons that fosters the participation of citizens, researchers, and educators in Europe.

\section{Methodology}
\label{sec:method}

We have explored and analysed current mainstream methods for designing use cases and evaluating common practice in digital cultural heritage use case design. 

To address the identified shortcomings, we propose a new methodology to advance the state-of-the-art in delivering use cases \cite{usecases}. We illustrate the methodology by providing five sample use cases which are key for reproducible research and collaborative analysis of digital texts, historical data, images, and digital twins.

\subsection{Use cases, user stories and scenarios}

Since use cases were suggested as a potentially useful tool in software technology development in 1986 by Ivar Jacobson~\cite{10.5555/993806}, the body of texts offering practical advice and methodologies for developing use cases has proliferated. Jacobson suggested that ``\emph{A use case is a special sequence of transactions, performed by a user and a system in a dialogue}'' \cite{10.1145/38765.38824}. The gradual movement towards federated and distributed eInfrastructures, where one system communicates with others, has added to the complexity of describing use cases as they move away from the interactions of a single user with a single system. The responses of a system with regard to other systems increase the complexity of use cases. This is also one of the major challenges in the ECCCH context.  

In software design literature, the terms user scenarios, use cases, and user stories are sometimes left undefined and used interchangeably. This lack of clarity is apparent also in the digital cultural heritage domain where ``use cases'' are often used as a term presenting various conceptualisations. To avoid confusion in terminology, within this work we are using the terms and definitions enumerated in Table \ref{tab:terms}.

\begin{table}[ht]
    \centering
        \caption{Definitions of terms used in this article}
    \label{tab:terms}
   {\small
    \begin{tabular}{p{0.1\textwidth}|p{0.35\textwidth}|p{0.5\textwidth}}
       \toprule
       \textbf{Term} & \textbf{Working definition} & \textbf{What is specific?} \\
       \midrule
       Use case & 
         (1) A methodology used to identify and organise system requirements.\newline
        (2) A set of sequences of interactions between systems and users in a particular environment and related to a particular goal. 
         &
    We will be using the term in the second sense. What characterises use cases is the use of short statements which are framed around specific actions. The order of these actions is essential. Use cases should also allow for different sequences of actions when using the system.\\ 
    \hline
    User story & Description of a system feature or a requirement told from the perspective of a user with a particular role. &  User stories are essential when information behaviours and needs are explored. A challenge with innovation in use cases is to identify user stories which are novel and bring value which has not been offered previously. This is an essential component of the preparatory work on the use cases where such user stories should be collected and analysed. \\
    \hline
    Use scenario & A detailed narrative describing the interaction of a user in a specific role with a system. A popular method to capture details about the roles is using personas. &
    Unlike the use cases, scenarios give a more generic feel of how the interaction of the user and the system happens. They are more linked to the contexts of use rather than being a succinct set of actions the user performs. \\
    \hline
    User &
    An actor (human or machine/software) who uses an infrastructure or a service. &
    The users in this work could be human end-users, e.g., researchers, or machine users, e.g., data providers or APIs. \\
    \bottomrule
    \end{tabular}
    }

\end{table}

The article does not have the ambition to offer a comprehensive analysis of the existing definitions. We provide further explanations on the way as we introduce and use the terms. 

\subsection{Previous work concerning use cases and digital cultural heritage}
Exploring the academic literature related to digital cultural heritage shows an impressive number of publications which discuss use cases (see Table \ref{tab:pubs}). It is not unusual that the term occurs more frequently in the context of Europeana, which has been developed since 2008, whereas publications directly related to ECCCH are still relatively rare. 

\begin{table}[ht]
    \centering
        \caption{Number of publications related to use cases and digital heritage (data retrieved in July 2025)}
    \label{tab:pubs}
    \begin{tabular}{p{6.5cm}|p{3.5cm}}
     \toprule
     \textbf{Search term} & \textbf{\# Occurrences\newline on Google Scholar} \\ 
     \midrule
     ``use case*'' and ``digital heritage'' & 2280 \\
   \rowcolor{Gray}
     ``use case*'' and ``digital cultural heritage'' & 1780 \\
    
     ``use case*'' and ``Europeana'' & 1590 \\
  \rowcolor{Gray}
     ``use case*'' and ``ECCCH'' & 8 \\
    \bottomrule
    \end{tabular}

\end{table}

However, exploring these publications shows that there is: 

\begin{itemize}
\item A vast variety of levels of detail and structures in the descriptions.
\item Multiple examples of publications where use cases are mentioned but actual descriptions of use cases are missing.
\item Scarce evidence on how the use cases have been implemented.
\item Limited insights relevant to eInfrastructures (demarcation of responses to systems with regard to other relevant systems in the digital cultural heritage domain). 
\item Unified Modelling Language (UML) and other visualisations, which are standard within the software development practice, are rarely used. Instead there is a tendency towards unstructured textual descriptions of use cases. 
\end{itemize}

Table \ref{tab:usecases} presents a snapshot of some more unconventional and insightful use cases, which appear in research publications and in projects developed in the last 10 years around digital cultural heritage. Our aim was not to build a complete inventory but rather to illustrate how diverse the areas could be in which use cases are being conceptualised. We should also mention that in some areas there was stronger evidence on previous systematic research on the developed use cases, e.g., in 3D digitalisation such an example is the work of Sander M\"unster~\cite{Munster2024}, while in other areas there are appearing use cases but no systematic analyses (e.g., in the domains of decolonisation and cost estimates of digital services). 

\begin{table}[ht]
    \centering
        \caption{Examples of innovative use cases based on CH content}
    \label{tab:usecases}
  {\small
    \begin{tabular}{p{0.1\textwidth}|p{0.25\textwidth}|p{0.5\textwidth}|c}
    \toprule
    \textbf{Domain} & \textbf{Strengths} & \textbf{Challenges} & \textbf{Source}  \\
    \midrule
    Data \newline aggregation & Comprehensive description in relation to various ecosystems &
    Use cases are framed as actions – these descriptions, although useful, are on a high level. & \cite{Freire2020Wikidata} \\
    \hline
    Biases\newline  in cultural heritage & Insightful observations of types of biases & What is called use cases are actually scenarios. & \cite{masschelein_2024_14514378} \\
    \hline
    Museum data & Sharing free, high quality images from the Rijksmuseum & Copyright restrictions. & \cite{Rijksmuseum_2013} \\
    \hline
    Digital cultural heritage & Democratise tools and break down long-standing barriers to cultural information & Biases, environmental impacts, and economic, social and cultural inequalities. & \cite{british-council-use-case} \\
    \hline
    Citation & Citation within the context of Institutional Repositories (IRs) & The article makes a case for use cases being essential within the IR context; we include this article as IRs are also part of the eInfrastructures which would contribute to ECCCH. However, the article provides neither a casual nor a formal description of the potential use cases. & \cite{vanderbosch} \\
    \hline
    Cost \newline Estimation of a \newline  Digital Service & A Pilot on Illuminated Manuscripts HUB & This is a casually described use case from E-RIHS. What makes this use case different is the described methodology for acquiring the use case -- interviews with domain experts. While this is a great starting point, it also could result in very institution-specific use cases rather than generic ones.  & \cite{deluca} \\
    \hline
    Structural granularity visualisation & Visualization of series’ categories and effective series in an archive  &  There is no formal use case description but the case is made that advanced digital tools for resource searching by keywords should also exploit semantics.  &\cite{demartino}  \\
    \hline
    Toolbox project as a use case & The RESTORE Data Integration Suite is a toolbox &  The article does not provide any formal description of the use case and one can argue either that this is a business level use case, or there is a complexity mismatch. & \cite{spadi}  \\
    \bottomrule
    \end{tabular}
    }

\end{table}

Other relevant efforts include the ongoing collection of user stories by the International Image Interoperability Framework (IIIF) initiative \cite{iiif} where 14 user stories had been documented. This is a useful example of a community effort to take stock of user stories in a specific domain.

\subsection{Defining use cases for JNs in federated eInfrastructures}
\begin{figure}[ht]
    \centering
    \includegraphics[width=0.5\linewidth]{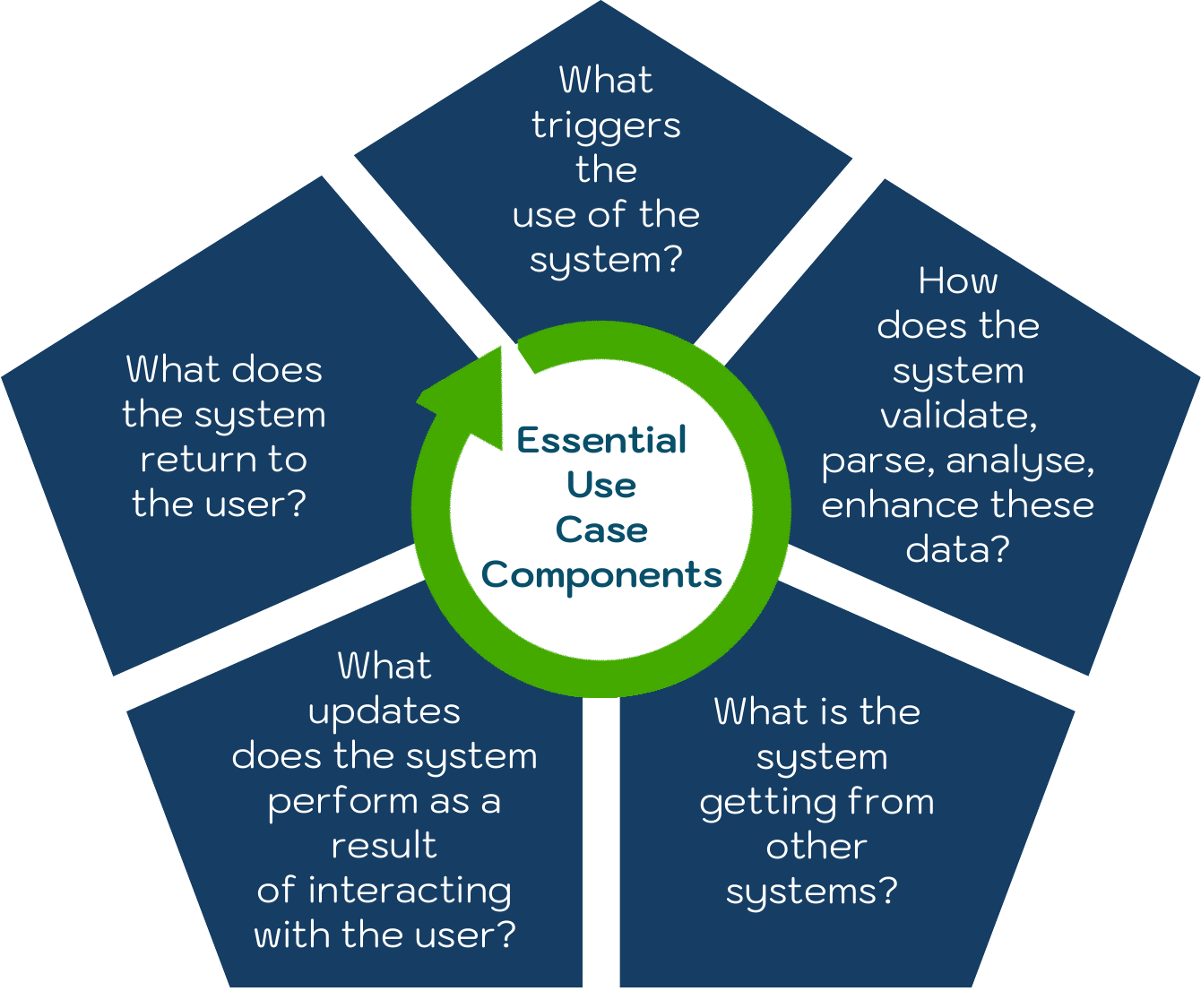}
    \caption{Five essential components to be captured in use cases}
    \Description{Five essential components to be captured in use cases}    
    \label{fig:five}
\end{figure}

In this article, we base our thinking on the recommendations of one of the Agile\footnote{\url{https://agilemanifesto.org/}} methodologies' pioneers, such as Alistair Cockburn \cite{Cockburn1}. According to his work, there are five essential components which need to be captured in use cases (Figure \ref{fig:five}): 

\begin{itemize}
    \item \emph{What triggers the use of the system} -- what are the actions undertaken by the user? What data is the user submitting? (this could be expressed via a user story)
    \item \emph{How does the system validate, parse, analyse, enhance these data?} 
    \item \emph{What is the system getting from other systems?} This is especially relevant to the ECCCH efforts. 
    \item \emph{What does the system return to the user} -- and what further actions of the users are triggered from the response? 
    \item \emph{What updates does the system perform as a result of the interaction with the user?}
\end{itemize}

The use cases can be described casually or formally \cite{Cockburn1}. Having in mind the current stage of development of use cases in digital heritage, our descriptions are \emph{casual}. Use cases are not expected to describe in detail data sources and components of the user interface. However, in our descriptions we mention a number of potential data sources. This is done with the intention not to dwell on specific data structures but to illustrate the diversity of systems which in fact are used in federated eInfrastructures. We also comment on the potential for ECCCH integration -- we picked this infrastructure for the purpose of offering specific examples but the use cases can be applied to other eInfrastructures. 

Our aim with these use cases is to illustrate several situations of answering different user needs, where the needs of a Digital Commons are taken into consideration. All use cases are focused on JNs enabling reproducibility and transparency. 

Following our analysis of the state of the art and current gaps, we identified five overarching themes that encompass areas of intensive development such as digital twins and AI, an area still underdeveloped in the Digital Humanities (e.g., High-Performance Computing Use (HPC)), and areas supporting sustainability and broader contributions to open research: 

\begin{itemize}
\item \textbf{Using and reusing JNs in HPC environments.} This use case seeks to enable faster and more efficient processing of complex tasks and collections, and large datasets.
\item \textbf{Application of JNs in digital twins.} This use case will demonstrate how data relevant to various digital twin models can be supported by JNs, which strengthens, in particular, the visualisations and re-use of digital twins. 
\item \textbf{AI preservation pilot use case.} This use case will explore how to approach the long-term preservation of AI models which are contributing to JNs. AI preservation in general is still a largely unaddressed area of work, despite its rapid integration into digital preservation tools and workflows. Its use across domains of research makes it critical to the reproducibility of results, and its proper long-term management an essential condition for the future robustness of ECCCH.
\item \textbf{Sustainability for JNs.} This use case will explore more broadly all the aspects contributing to uninterrupted future use of JNs, including an audit of FAIRness of JNs (delivering findable, accessible, interoperable and reusable JNs).
\item \textbf{Open Science for JNs.} This use case will address Open Science for data and code sharing using platforms such as Zenodo and making the entire research process more transparent and accessible, promoting collaboration, and maximizing the impact of research. 
\end{itemize}
    
By developing these themes, we offer the following novel components: 

\begin{itemize}
\item \textbf{\emph{Rich data fabric:}} For all themes, we will develop use cases that incorporate the researcher’s personal context and tasks, triggering communication with ECCCH. It will pull data from the ECCCH trio of major infrastructures, including the ECCCH itself, DS4CH, and TMO, as well as external open resources like Wikidata and GLAM collections accessible via institutional websites.
\item \textbf{\emph{Robust use case methodology:}} We are basing our use case development on a methodology which takes as a basis the best practice in software development that includes attention to flexibility to adapt to changes and faster delivery of working software, as well as incorporating components specific to the digital cultural heritage domain. We have the ambition to extend this methodology to be applicable across domains. 
\item \textbf{\emph{Easy integration into other ECCCH efforts:}} We will be working alongside previously funded ECCCH-related projects to demonstrate how our use cases can be applicable in their specific subject domains (AUTOMATA,\footnote{\url{https://www.echoes-eccch.eu/AUTOMATA/}} TEXTaiLES,\footnote{\url{https://www.echoes-eccch.eu/textailes/}} HERITALISE\footnote{\url{https://www.echoes-eccch.eu/HERITALISE/}} and those still under negotiation). 
\item \textbf{\emph{Collaboration with other EU infrastructures:}} We will collaborate with the DS4CH and integrate cloud resources provided by ESOC, such as the Interactive Notebooks service \cite{eosc-jns}.\footnote{\url{https://open-science-cloud.ec.europa.eu/services/interactive-notebooks}} 
\end{itemize}

\section{Five examples of JN-focused use cases for federated eInfrastructures}
\label{sec:use-cases}
The methodology proposed in Section \ref{sec:method} follows a shared approach which will be applied among five thematic areas, which illustrate different user needs and system collaboration. 

\begin{figure}[ht]
    \centering
    \includegraphics[width=\linewidth]{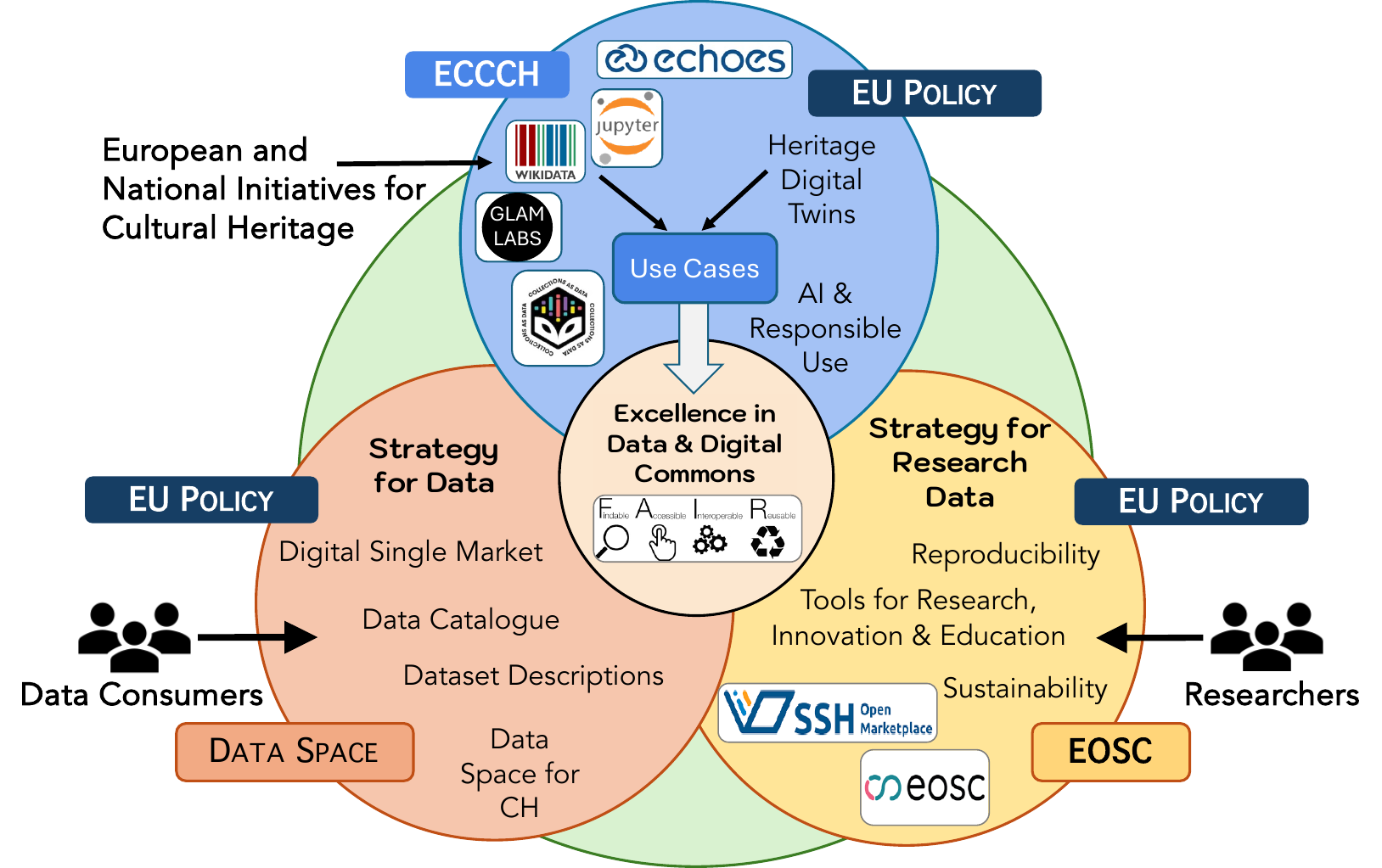}
    \caption{Integration of existing platforms and services to create the use cases in the ECCCH, to foster excellence in data and Digital Commons. Note that this figure was inspired by the ECHOES project website and tries to illustrate how our approach fits within the current research data infrastructure}
    \Description{Integration of existing platforms and services to create the use cases in the ECCCH, to foster excellence in data and Digital Commons}
    \label{fig:arquitecture}
\end{figure}

Figure \ref{fig:arquitecture} shows how existing European infrastructures are connected and collaborate to create the use cases proposed in this work and in the context of the ECCCH, fostering excellence in the creation of a Digital Commons.

\begin{figure}[ht]
    \centering
\includegraphics[width=0.7\linewidth]{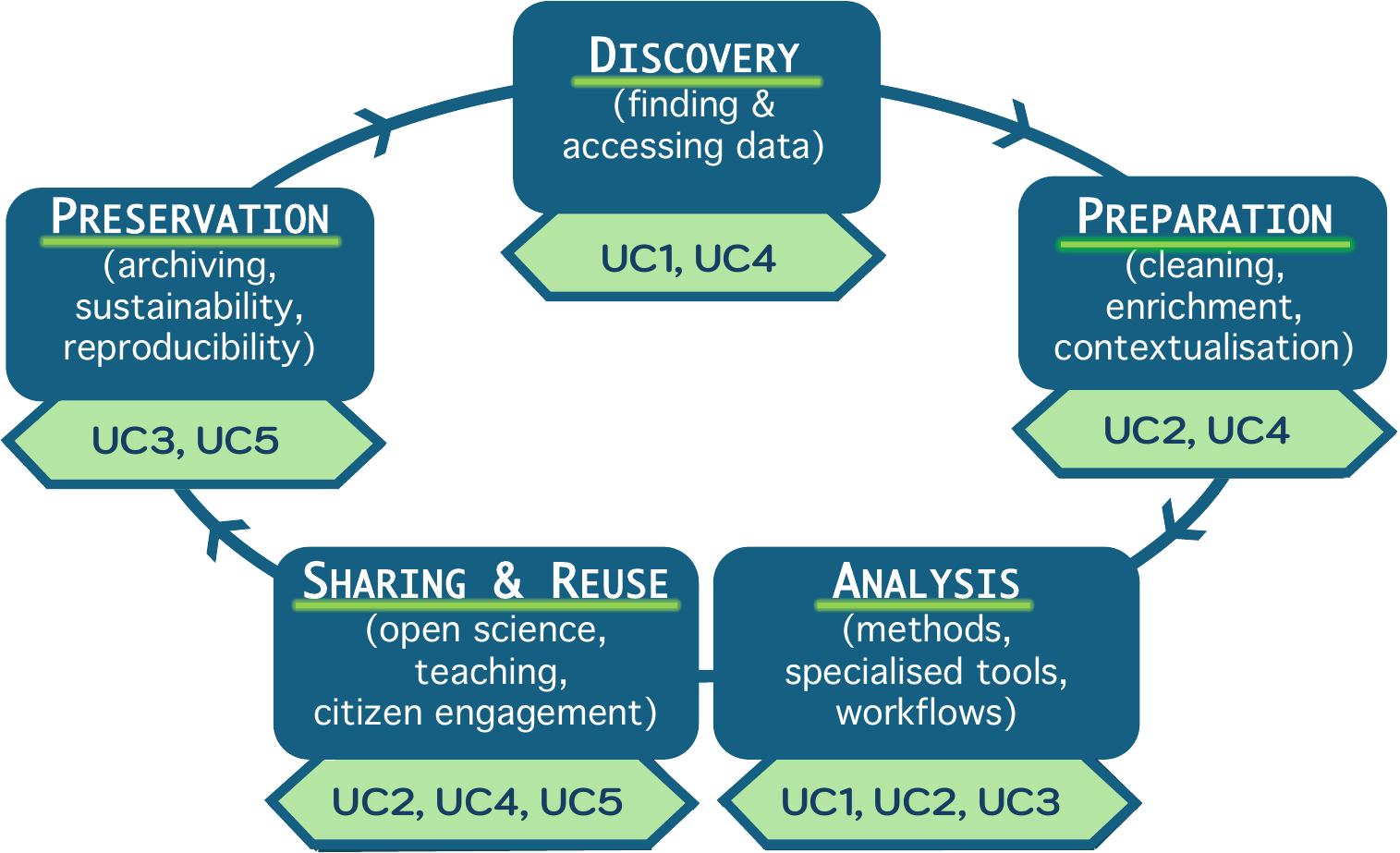}
    \caption{The research lifecycle is divided into a number of tasks. Each of the use cases proposed in this work focuses on some of these tasks}
    \Description{The research lifecycle is divided into a number of tasks. Each of the use cases proposed in this work focuses on some of these tasks}
    \label{fig:lifecycle}
\end{figure}

The implementation of the use cases will employ a set of Key Performance Indicators (KPIs) to track progress, measure success, and make informed decisions. Some examples from various areas such as performance and software development include: number of JNs created, number of datasets reused, user experience, code quality metrics, resilience and security. Figure \ref{fig:lifecycle} shows how the research lifecycle divides the implementation of use cases into a discrete number of tasks including: discovery, preparation, analysis, sharing and reuse, and preservation.

% replaced this because red on green doesn't work for colour blind readers

The following sections describe the use cases proposed in this work as a result of the application of the methodology.

\subsection{Using and reusing JNs in high-performance computing environments}

GLAM institutions host and provide access to historical maps that depict past interpretations of reality and events, reflect cultural and social trends of the time, offer artistic value, and track the evolution of cartography. For example, (i) KU Leuven Libraries hosts more than 20,000 maps and atlases with a focus on Belgian territory from the 16th century to the present; (ii) the Royal Danish Library map collection contains the largest total documentation of Danish cartography and map production, and it is also one of Northern Europe's largest collections of maps; and (iii) additional institutions and aggregators such as Europeana and the national libraries of Spain and Scotland provides access to historical maps covering different topics and periods. These rich materials show researchers and European citizens how the landscape used to look, the continent's rich and complex past, and thus offer essential information for geographers and historians. 

The digital collections are made available in a wide diversity of formats such as Zip files and by means of an API, enabling access and computational use. However, the digital collections are often published as individual sets, hindering collaboration and integration with other resources. Combining data from various sources allows for a more comprehensive and complete view of the content, leading to new and possibly novel insights. The integration of several datasets can be performed in several ways, such as the creation of links between the resources to create a knowledge graph. LLM, in combination with traditional Entity Linking methods, enhance accuracy and efficiency when connecting the terms or entities mentioned within the content to unique, well-defined corresponding entities.

For centuries, maps have played a predominant role in modern Western society, being a valuable resource in strategic fields, such as astronomy, geometry, printing, geodetic measurement systems, optics, as well as aerospace technology, computer and information science \cite{maps}. This use case intends to answer the following questions: (1) \emph{To what extent are historical maps visually useful to demonstrate how territories, borders, and place names have changed over centuries in Europe?} (2) \emph{How different cultures influenced each other in the context of trade and exploration?} (3) \emph{In education, how do maps represent different interpretations of the past?}

This use case integrates the use of several collections, moving beyond data silos and demonstrating the benefits and advantages of using collections together with other resources. The use case applies predictors for the type of textual content on maps representing buildings, icons, mountains, etc. and linking it to gazetteers, indexes of places and related metadata to facilitate finding and interpreting maps \cite{mahowald2024integratingvisualtextualinputs}. Map data can later be used as training data for future Geographic Information Systems (GIS), AI and ML tools for automatic map understanding. It will also promote the use of map text as data within communities of humanities researchers, the sciences and the cultural heritage sector, creating awareness and developing capacity for further formulation and implementation of policies to support the sharing of important map collections as data.

This use case also demonstrates the potential of existing cloud and HPC infrastructures in Europe in combination with JNs, LLM and AI models to apply computer vision, identify relevant entities, and generate new metadata to facilitate new ways of browsing the data. It incorporates a generalisable and reproducible ML pipeline based on JNs to process and reuse metadata and text on historical maps. These JNs can be run and hosted in the ECCCH enabling security and sustainability and avoiding common issues concerning the publication and future reuse of JNs such as data leakage, dependency issues, and security breaches. 

Potential users and audience for the use case includes researchers, scholars and lectures from different disciplines such as historians, humanities, science and geographers, practitioners and students, will be able to run and customize the reproducible code to adapt it to their own needs and requirements as well as learn and contribute to reproducible and open research by using the JNs as training materials; librarians, curators and editors of historical map digital collections; Cultural Heritage and Cultural and Creative Industries (CCIs) related to the use of historical maps; and students at high school and university degrees.

\textbf{Relevance}. Integrating the data from several institutions into the ECCCH can facilitate best practices and guidelines for other institutions willing to start using these services. The collections will be enriched with additional detailed descriptions of the datasets by means of datasheets and machine-readable metadata facilitating the connection with other initiatives such as the DS4CH. The data generated, alongside reproducible examples for reuse and a pipeline based on JNs, will be made openly available for researchers and citizens. New uses of these outputs include the creation of training materials and workshops to promote their reuse. It will demonstrate how existing JNs services in EOSC can be combined and integrated into the ECCCH.

This use case describes a case of how researchers in any discipline interact with historical maps, making a significant contribution to European digitised historical data and understanding historical change and cultural and social context.

\subsection{Application of JNs in digital twins}
Digital twins  create  a virtual  replica  of  a  physical  product,  process  or system. They can be extremely useful to understand the impact of hypotheses and different scenarios for the reuse and integration of heritage to respond to contemporary societal challenges. They have been applied to different domains such as healthcare, manufacturing to simulate product performance, aerospace engineering and cultural heritage. Digital twins provide a powerful tool for simulating, analysing, and optimising a wide range of real-world systems and processes, leading to improved efficiency, reduced costs, and better outcomes. Digital twins offer a transformative and innovative technology with significant potential to improve various aspects of European society and economy, from industrial processes to environmental sustainability.

The European Commission is actively promoting the use of digital twins through initiatives like Destination Earth and the European Virtual Human Twins Initiative, highlighting their potential to address key societal challenges and drive innovation. Several laboratories have been created to experiment with digital twins such as the Virtual Environments Lab at the Cyprus Institute and the Time Machine Organisation. However, despite all these efforts, many challenges still remain such as interoperability, standardisation or a skills gap between staff working in private companies and those in publicly funded institutions.

This use case will demonstrate how data relevant to various digital twin models can be supported by JNs, which strengthens, in particular, the visualisations and re-use of digital twins. It will explore how JNs can enhance interoperability for digital twins by providing metadata, data and paradata descriptions \cite{Ioannides2025}. By using existing services in EOSC concerning the use of JNs, the DS4CH and data curation and enrichment platforms, this use case will explore digital twins in a reproducible environment to illustrate how the data can be reused in innovative and creative ways.

The audience includes creative studios and agencies who use digital technologies and cultural heritage assets, e.g., digital twins for business products or services; GLAMs of all sizes (national, regional, local, independent, site museums and special collections) who are commissioning, creating and using digital twins; professionals in the private sector with a special focus on those with an interest in the cultural heritage; researchers and lecturers in different disciplines; and individuals who create, use and reuse digital twins.

\textbf{Relevance}. This use case will demonstrate how data relevant to various digital twins models can be supported by JNs to provide examples of use. Existing networks such as the Local Time Machines -- as a network of 85 regional and local networks, can help as co-creators and users of this use case. This use case will enhance how researchers interact with digital twins, making a significant contribution to European digitised data.

\subsection{AI preservation use case}

Advances in technology have paved the way to a new context in which AI has become increasingly important in today's world, as it has the potential to revolutionise many industries, including healthcare, finance, education, and more. The use of AI has already improved efficiency, reduced costs, and increased accuracy in various fields. The EU’s approach to AI is based on excellence and trust, with the aim of boosting research and industrial capacity while protecting public safety and fundamental human rights. AI provides different tools and services that affect the way knowledge is produced today. In order to understand how AI is advancing society forward, our society needs access to specific versions of these tools and services, and researchers in particular should be able to reproduce experiments which rely on complex, computer-generated logic and analyses.  

Meanwhile, user-friendly generative AI technologies are already quite sophisticated and evolving so fast, it is a challenge for contemporary researchers to share prompts and models which reliably generate the same results let alone preserve them with adequate documentation for long-term reuse. There is an urgent need to define the significant properties of AI to effectively preserve them for future generations~\cite{Broussard}. Existing data storage systems can be affected by known phenomena such as \emph{bit rot}, resulting in data integrity issues due to several reasons such as physical medium breakdown and magnetic decay, leading to inaccessible, inaccurate, or lost files. Common and traditional preservation practices include storing distributed copies of data, using, if possible, open formats~\cite{da-eu}. However, AI preservation will require the consideration of AI models, training data and infrastructure and dependencies such as PyTorch, all crucial elements to assess AI performance \cite{PAdilla_AI}.

During the last few years, there has been modest work concerning AI preservation, which has been mainly focused on AI applications to the fields of digital preservation and heritage science. This use case will explore how AI preservation can be applied to the research datasets and tools in the ECCCH in order to foster a more sustainable and reproducible use of AI. Some questions to be answered include, for example: (1) \emph{Should provenance documentation and metadata about training data be preserved instead of the training data itself?} (2) \emph{What volume of data should be preserved to ensure and facilitate long-term reuse and where is this data best stored -- especially within federated eInfrastructures?} (3) \emph{How should data creators document the use of AI in their work, in light of the constant evolution of training datasets and models?} (4) \emph{What are the issues and challenges concerning the environmental impact of AI preservation?}

This use case will explore AI preservation in the context of the ECCCH by defining a reproducible pipeline using JNs. Existing initiatives concerning AI preservation will be reviewed and analysed in order to be adapted to the ECCCH. Some examples include Model Cards from Hugging Face, datasheets for cultural heritage datasets and existing workflows to publish and reuse GLAM collections as data. Current developments in research data appraisal, curation and preservation management, especially in the European context with the EOSC EDEN and FIDELIS projects,\footnote{https://eden-fidelis.eu} will also inform this work. The pipeline will be applied to a selection of datasets provided by relevant institutions such as Europeana, the Digital Repository of Ireland, KU Leuven and the National Library of Scotland to identify best practices and provide guidelines for CH institutions. It will be of use particularly to the development of institutional position statements on AI in cultural heritage data reuse, which are only beginning to emerge from the sector. This use case will pave the way for similar institutions willing to participate in the ECCCH and to harness the long-term preservation of AI models that contribute to JNs. 

The audience for this use case consists of GLAMs of all sizes (national, regional, local, independent, site museums and special collections); policy-makers including governmental bodies involved in designing and promoting preservation policies on a European and national level; NGOs and associations; higher education institutions, educators and students engaged in formal post-secondary education; users of Europeana collections.

\textbf{Relevance}. AI preservation is becoming crucial for the reproducibility of research studies and, more generally, for society to understand how knowledge is produced today. The datasets and code informing this use case will be enhanced with detailed provenance documentation and integrated into the ECCCH. Training materials will be made available in order to foster its adoption by researchers and its successful application by cultural heritage and digital preservation professionals. 

Aligned with the European AI Strategy \cite{eu-ai} as well as national initiatives to address the use and application of AI, this use case will help to better understand the need and requirements for long-term AI preservation and make the EU a world-class hub for AI, building an environment in which research remains human-centric, trustworthy and responsible. As society continues to embrace AI, it is crucial that we remain mindful of its impact and work to address the challenges that come with its evolution. AI preservation ensures that AI contributes to the global vision of research data as a public good, capable of improving people's lives as well as creating a better future for generations to come.

\subsection{Open Science for JNs}
The European Digital Infrastructure Consortium (EDIC)\footnote{\url{https://digital-strategy.ec.europa.eu/en/policies/edic}} ``Digital Commons'' initiative aims to strengthen digital autonomy and sovereignty in Europe. Digital Commons are resources, like open-source software, open data, or open standards, that can be used by multiple people simultaneously. Digital Commons serves as the foundation for Open Science which is at the centre of European research policy, enabling increased collaboration, faster research dissemination, enhanced reproducibility, and greater public engagement. Notable Open Science practices include: (i) data deposition in shared repositories in line with the FAIR principles; (ii) providing open access to scientific publications, research data, models, algorithms, software, notebooks, workflows, and all other research outputs; (iii) ensuring verifiability and reproducibility of research outputs; and (iv) promoting public engagement in research and innovation, bolstering citizen science and enhancing public trust in science. Digital Commons and Open Science are interconnected concepts that promote the sharing and accessibility of knowledge. The synergy between these two concepts is crucial for advancing scientific progress and fostering a more open and inclusive research ecosystem.

GLAM and CH institutions host rich content in the form of digital collections. While they have followed best practices and guidelines to make them available for the public, there are still challenges to be addressed in order to consolidate a complete Open Science and Digital Commons approach. Some examples include data deposition in research repositories, the publication of examples of use, the description of the dataset by means of meaningful documentation describing data provenance and quality in detail or the dissemination and engagement with the public to maximise the impact. JNs have emerged as an interactive and versatile tool that transforms how researchers, developers, institutions and data scientists conduct and communicate their work. GLAM and CH institutions have started to use them to show how to reuse their digital collections providing examples of reuse and access to their content. 

According to the current situation, this use case intends to answer the following questions: (1) \emph{How can Open Science and Digital Commons be addressed by means of JNs in the ECCCH?} (2) \emph{How can we showcase, extend and improve the ECCCH to fully support Open Science and Digital Commons in order to engage with the public?}  

This use case will create a reproducible and executable workflow to integrate, publish, reuse and disseminate digital collections. It will be built upon a selection of openly available digital repositories providing a wide diversity of content such as scholarly communications, bibliographic metadata and CH data. The data will be integrated to create a unified and comprehensive view, unlocking its potential for analysis and avoiding data silos. Examples of use will be implemented such as AI-based discovery tools using emergent techniques such as Retrieval Augmented Generation (RAG). This use case will also demonstrate how federated queries can be employed to integrate and search across datasets. It will also use Zenodo’s API to integrate the outputs and enable an efficient reuse of the data.

The audience includes GLAMs of all sizes (national, regional, local, independent, site museums and special collections); professionals working with GLAM sector institutions who create content, promote culture, curate exhibitions, manage collections and develop educational programmes; policy-makers, including governmental bodies, are involved in designing and promoting cultural policies on a European and national level; cultural and heritage NGOs and associations who help raise awareness of cultural resources and communities through their online portals and networks; higher education students and users of Europeana digital collections and metadata. 

\textbf{Relevance}. Aligned with the Digital Commons and Open Science initiatives, this use case aims to strengthen digital autonomy and sovereignty in Europe, providing best practices and easy-to-follow guidelines. It will enable the humanities, scientific, and cultural heritage communities to adopt Open Science and Digital Commons to increase the visibility of the digital collections by ensuring verifiability and reproducibility of research outputs.

This use case showcases how existing data provided by relevant European institutions can be integrated avoiding data silos and providing new insights. It identifies best practices in how the data generated can be employed for AI in terms of data quality and performance.

\subsection{Sustainability for JNs}
JNs have recently emerged as a powerful tool that lowers many barriers concerning reproducibility in Open Science. JNs were designed to support reproducible research by enabling scientists to easily integrate computational narratives combining text, code, charts and documentation. Aligned with the ECCCH principles, JNs promote inclusion and democratise open and transparent access to knowledge and digital collections, which are understood as public goods by and for all. Researchers are citing their JNs in their publications, and they can be shared with the community, enabling collaboration, and the reproducibility and verification of results. In research in the humanities, there is a growing interest by researchers and expanding supply by cultural heritage institutions. JNs are a core avenue for use, enrichment and integration of cultural heritage digital collections in the form of single, shareable documents. However, a focused effort on harnessing the potential of JNs within ECCCH is still lacking which marginalises the Digital Humanities research need of using multiple collections as data and other open digital resources such as Wikidata, exemplified by JNs. 

This use case focuses on the aspects contributing to uninterrupted future use of JNs, including an audit of the FAIRness of JNs. This use case covers the following questions: (1) \emph{What are the current challenges for future reuse of JNs? How can the ECCCH provide sustainability for JNs?} (2) \emph{According to previous work concerning FAIR JNs, what are the refinements needed to be successfully adapted for the ECCCH?}

Built upon previous work with regard to international initiatives to promote FAIR JNs and computational access in GLAM such as the International GLAM Labs Community, Library of Carpentry{\footnote{\url{https://librarycarpentry.org/}}} or Australia Research Data Commons,\footnote{\url{https://ardc.edu.au/resource/fair-for-jupyter-notebooks-a-practical-guide/}} this use case will address the aspects of reproducibility, sustainability, licence and provenance~\cite{DBLP:journals/jasis/CandelaCS23}. These characteristics are essential to promote sustainable use of JNs in the European context. 

The audience includes SMEs (Small and Medium-sized Enterprises/businesses) operating in CH; professionals and freelancers operating Cultural Heritage and Cultural and Creative Industries (i.e., tourist guides, tour operators, etc.); national policy-makers, governmental bodies and associations that are operating within digital content and CH industries; museum curators or other culture based staff involved in CH sector; service providers for cultural and creative industries players.

\textbf{Relevance}. Aligned with the Digital Commons and Open Science initiatives, this use case will explore sustainability in JNs to enable future reuse. It will facilitate the humanities, scientific, and cultural heritage communities best practices and guidelines to make their JNs aligned with FAIR. This use case will showcase how existing JNs projects provided by relevant European institutions can be assessed in terms of sustainability. While there are currently JN services in the European Open Science Cloud, challenges still remain concerning the sustainability and future use of JNs. This use case will explore the current situation and provide best practices to overcome these issues and make Europe a leader in terms of reproducible research. 

\subsection{Limitations and potential for future work}

This work has limitations in terms of the current development of the research data infrastructure, the data employed, scope, and coverage. Some examples are described below.

Despite ongoing efforts to provide a collaborative and efficient data research infrastructure in Europe, such as ECCCH and DS4CH, challenges remain. Some examples are: (i) the correct definition of uses cases in order to provide an easy-to-adopt approach as well as enable their extension; (ii) enabling researchers to reproduce their research results in a sustainable and integrated way; or (iii) integrating existing data to avoid data silos. There is a need to promote these initiatives among small and medium institutions as well as CCIs.

The use cases provided as a result of this work have been defined according to the same methodology following the same stages of identifying, extracting and enhancing data. They can be compared according to different aspects, enabling their analysis. For instance, they have similarities according to how they are co-created by the community, the narrative and structure, goals in terms of Digital Commons and Open Science, and integration with the ECCCH. However, there are also differences which relate to aspects of content reuse (e.g, maps, AI models and digital twins) and innovation covering relevant aspects such as AI preservation and digital twins. The use cases also illustrate processes which can be used in different stages of the research lifecycle. In addition, they all are aligned in terms of sustainability and explore the connection with other notebook initiatives such as Google \footnote{\url{https://colab.research.google.com}} and Kaggle Notebooks.\footnote{\url{https://www.kaggle.com/code}}

We discussed the need for a shared methodology in the development of use cases within the domain of digital cultural heritage. Our article includes examples that are of the kind of casual descriptions or even precursors of proper formal use cases. A further conceptualisation effort would transform the use cases into succinct lists of specific steps. Before proceeding in this direction, we wanted to share our proposals and discuss with the wider digital heritage community the suitability of this approach. A brief description of the audience and relevance is provided for each use case to highlight the benefits of their implementation in the current European data infrastructure context. Note that the ECCCH and DS4CH are currently under development. This work may need additional adaptations and refinements once these projects are finalised in the coming years.

Concerning the literature review, we have used an approach of picking a range of examples, which helped us to formulate challenges in the current stage of digital heritage eInfrastructures. While we have illustrated that there are thousands of publications which mention use cases but in fact do not discuss them in detail, there are also publications which do not mention use cases but formulate useful user stories. This requires further research. We plan to dedicate time in the future to selecting a knowledge organisation system that will be suitable for creating in the future a long list of use cases and user stories from the body of scientific publications and ongoing projects. 

An additional question which has not been explored in this article is with which system exactly the user starts their communication. The assumption is that the user starts by using ECCCH or another infrastructure, but they could also start querying an institutional repository, which then requests further data for enrichment from ECCCH. This work also form the basis for future work around other types of notebooks such as Observable.\footnote{\url{https://observablehq.com}} This research would be a useful to provide guidelines to create training materials (e.g., set of templates) to write consistent use cases that follow a specific framework. This requires further analysis and depends on the architecture that ECCCH would gradually adopt and deliver. 

\section{Implementing the use cases: a workflow approach}

According to previous work and best practices concerning the definition of best practices and workflows to publish and reuse CH data in research \cite{DBLP:journals/jasis/Lee25,10.1108/GKMC-06-2023-0195,transformations:14729}, Figure \ref{fig:workflow} shows a workflow describing the main steps to implement the use cases proposed in this work. 
This workflow could be implemented using open source tools such as Python and Apache Airflow.\footnote{\url{https://airflow.apache.org}} Each of the steps is briefly described below:

\begin{figure}[ht]
    \centering
\includegraphics[width=\linewidth]{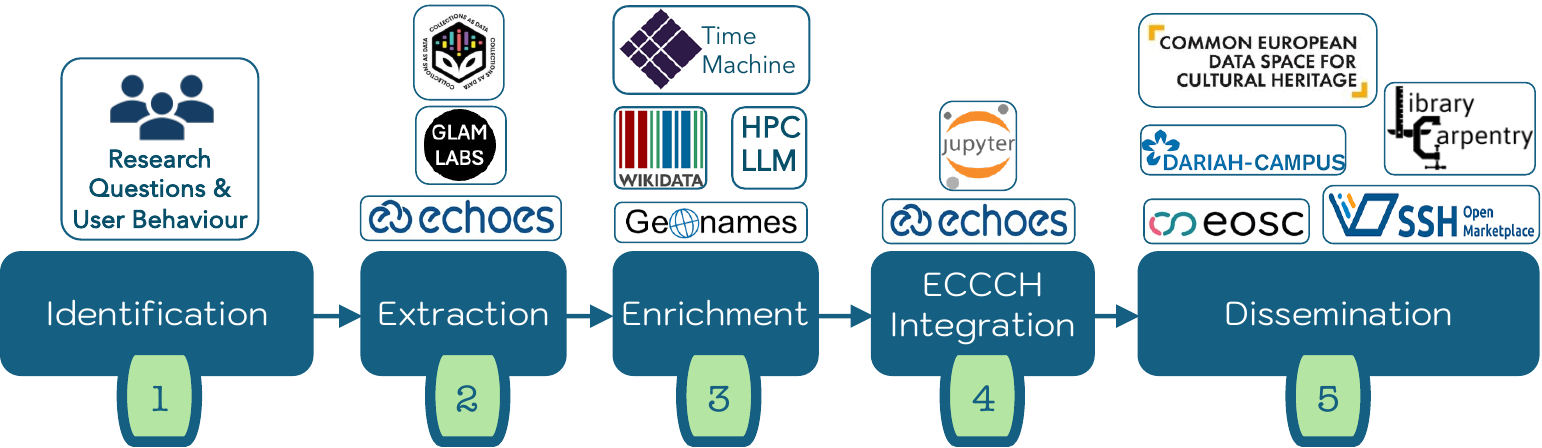}
    \caption{Workflow to implement the use cases proposed in this work}
    \Description{Workflow to implement the use cases proposed in this work}
    \label{fig:workflow}
\end{figure}

\textbf{1. Identification.} The first step corresponds to the identification of the datasets to be used to answer a research question. For example, how can relevant people such as artists related to the Spanish Republic and exile be analysed by means of reproducible methods? This step also includes the identification of the digital collections and data required to perform the analysis. GLAM Labs, such as the Data Foundry at the National Library of Scotland,\footnote{\url{https://data.nls.uk/}} and research repositories provide digital collections in the form of data and machine-actionable collections enabling computational use. Other options may include GitHub repositories or data platforms such as Zenodo.

\textbf{2. Extraction.} This step involves the extraction of the data. The institutions have followed a wide diversity of approaches to make the digital collections available. As a result, this step can follow different approaches such as downloading a Zip file or using an Application Programming Interface (API). The data is available in different formats such as JSON or XML. In some cases, the data can be transformed to other formats such as Resource Description Framework (RDF) to make it more usable and accessible \cite{w3c-rdf}.

\textbf{3. Enrichment.} The extracted data will be enriched with external repositories such as Wikidata and GeoNames using innovative methods such as entity linking and LLM. This will provide a more integrated approach avoiding data silos and enabling transnational collaboration between institutions and countries. This will facilitate a federated approach to access and reuse the content by means of techniques such as SPARQL~\cite{w3c-sparql}.

\textbf{4. ECCCH integration.} This step involves the integration with the ECCCH in terms of data publication, promotion and reuse. The final data sets will be described according to the ECCCH metadata model ensuring high quality and detailed metadata. Narrative and machine-readable documentation will be provided according to best practices promoted by the community such as the datasheets \cite{Alkemade-2023}. The European Open Science Cloud  EU Node supports multi-disciplinary and multi-national research promoting the use of FAIR (Findable, Accessible, Interoperable, Reusable) data and supplementary services such as JNs and large file transfers. Combining these data and cloud services, a collection of JNs illustrating how to reuse the extracted data including open and reproducible code runnable in the cloud will be included as part of the use case \cite{Candela-2025,transformations:14729}. This will enable the reproducibility of all of the steps in order to adopt Open Science and collaboration as main principles.

\textbf{5. Dissemination.} The results will be disseminated to be widely used by European cultural heritage professionals and researchers, enabling new ways to interact, cooperate and co-create, thus supporting the generation of new knowledge and opening of new research paradigms. Several platforms will be used such as the Social Sciences and Humanities Open Marketplace\footnote{\url{https://marketplace.sshopencloud.eu}} and the European Open Science Cloud.

\section{Conclusion}

There is a growing interest in the role that research data infrastructures play in the reproducibility of research and the sustainability of data sharing practices across sectors. Several European initiatives, such as EOSC, DS4CH and ECCCH, are positioned to take advantage of the digital shift in the cultural heritage sector to increase access to culture as a computational resource, enabling the humanities, sciences, and broader cultural heritage sector to engage fully in Open Science practices and cross-disciplinary exchange. These infrastructures also create the potential for the integration and reuse of data siloed in existing scientific repositories, fostering responsible AI-driven discovery and strengthening digital autonomy and sovereignty in Europe.

Based on previous work, this contribution provides a methodology for defining and implementing use cases for the ECCCH. These contributions are intended to leverage the adoption and use of the ECCCH by GLAM institutions and CCIs, providing detailed documentation, training materials and reproducible code.

Future work to be explored includes refining and implementing the workflow to create the use cases. In addition, the application and adaptation of the methodology to other domains, such as the protection of the environment and healthcare, will have to be explored.

\begin{acks}
We would like to thank the International GLAM Labs Community for their support and guidance in this work.
\end{acks}

\bibliographystyle{ACM-Reference-Format}
\bibliography{biblio}

\appendix
\section{Glossary}

    \begin{tabular}{r l}
    \toprule
\textbf{AI}& Artificial Intelligence\\ \rowcolor{Gray}
\textbf{AI4LAM}& AI for Libraries, Archives, and Museums \\
\textbf{API}& Application Programming Interface\\ \rowcolor{Gray}
\textbf{CCIs}& Cultural Heritage and Cultural and
Creative Industries\\
\textbf{CH}& Cultural Heritage\\ \rowcolor{Gray}
\textbf{DS4CH}& European data space for cultural
heritage\\
\textbf{ECCCH}& European Cloud for Heritage Open Science\\ \rowcolor{Gray}
\textbf{ECHOES}& European Cloud for Heritage Open Science Project\\
\textbf{EDIC}& European Digital Infrastructure Consortium\\ \rowcolor{Gray}
\textbf{ENA}& Europeana Network Association\\
\textbf{FAIR}& Findable, Accessible, Interoperable and Reusable\\ \rowcolor{Gray}
\textbf{GLAM}& Galleries, Libraries, Archives, and Museums\\
\textbf{HPC}& High-Performance Computing \\ \rowcolor{Gray}
\textbf{IIIF}& International Image Interoperability
Framework\\ 
\textbf{IRs}& Institutional Repositories\\ \rowcolor{Gray}
\textbf{JNs}& Jupyter Notebooks\\
\textbf{KPIs}& Key Performance Indicators\\ \rowcolor{Gray}
\textbf{LLM}& Large Language Models\\ 
\textbf{ML}& Machine Learning\\ \rowcolor{Gray}
\textbf{RAG}& Retrieval Augmented Generation\\
\textbf{RDF}& Resource Description Framework\\ \rowcolor{Gray}
\textbf{SME}& Small and Medium-sized Enterprise/Business\\
\textbf{TMO}& Time Machine Organisation \\ \rowcolor{Gray}
\textbf{UML}& Unified Modelling Language\\
\textbf{UKRI}& UK Research and Innovation\\
    \bottomrule
    \end{tabular}
\end{document}